\documentclass{aastex}
\usepackage{emulateapj5}
\shorttitle{A Chandra view of NGC 3621}
\shortauthors{Gliozzi et al.}

\newcounter{species} 
\def\ion#1#2{\setcounter{species}{#2}{\rm #1}$\,${\rm\scriptsize\Roman{species}}\relax}

  \def\ngc{NGC~3621}
  \def\feka{Fe K$\alpha$}
  \def\chandra{{\it Chandra}} 
  \def\xmm{{\it XMM-Newton}}
  \def\spitzer{{\it Spitzer}} 
   
  \def\hst{{\it HST}}

  \def\lum{erg s$^{-1}$}
  \def\flux{erg cm$^{-2}$ s$^{-1}$}
  \def\nh{cm$^{-2}$}
  \def\arcsec{$^{\prime\prime}$}

  \def\ltsima{$\; \buildrel < \over \sim \;$}
  \def\simlt{\lower.5ex\hbox{\ltsima}} 
  \def\gtsima{$\; \buildrel > \over \sim \;$}
  \def\simgt{\lower.5ex\hbox{\gtsima}} 

\begin{document}
\title{A Chandra view of  NGC 3621:\\
a bulgeless galaxy hosting an AGN in its early phase?}

\author{Mario Gliozzi}
\affil{George Mason University, 4400 University Drive, Fairfax, VA 22030}

\author{Shobita Satyapal}
\affil{George Mason University, 4400 University Drive, Fairfax, VA 22030}

\author{Michael Eracleous\altaffilmark{1}}
\affil{Department of Astronomy \& Astrophysics and Center for Gravitational Wave Physics, The Pennsylvania State University, 525 Davey Lab, University Park, PA 16802}
\altaffiltext{1}{Center for Gravitational Wave Physics, The Pennsylvania State University, University Park, PA 16802}

\author{Lev Titarchuk}
\affil{NASA Goddard Space Flight Center, Code 662, Greenbelt, MD 20771}
\affil{George Mason University, 4400 University Drive, Fairfax, VA 22030}

\author{Chi C. Cheung}
\affil{Astrophysics Science Division, NASA Goddard Space Flight Center, 
Greenbelt, MD 20771}

\begin{abstract}
We report the detection of a weak X-ray point source coincident with
the nucleus of the bulgeless disk galaxy NGC 3621, recently discovered
by \spitzer\ to display high ionization mid-infrared lines typically
associated with AGN. These \chandra\ observations provide 
confirmation for the presence of an AGN in this galaxy, adding to the
growing evidence that black holes do form and grow in isolated
bulgeless disk galaxies.  Although the low signal-to-noise ratio
of the X-ray spectrum prevents us from carrying out a detailed
spectral analysis of the nuclear source, the X-ray results, combined
with the IR and optical spectroscopic results, suggests that NGC 3621
harbors a heavily absorbed AGN, with a supermassive black hole (SMBH)
of relatively small mass accreting at a high rate.  \chandra\ also
reveals the presence of two bright sources straddling the nucleus
located almost symmetrically at 20\arcsec\ from the center.  Both
sources have 0.5--8 keV spectra that are well-fitted by an absorbed
power-law model. Assuming they are at the distance of \ngc, these two
sources have luminosities of the order of $10^{39}$ \lum, which make
them ultra-luminous X-ray sources (ULXs) and suggest that they are
black hole systems. Estimates of the black hole mass based on the
X-ray spectral analysis and scaling laws of black hole systems suggest
that the 2 bright sources might be intermediate mass black holes with
$M_{\rm BH}$ of the order of a few thousand solar masses. However,
higher quality X-ray data combined with multi-wavelength observations
are necessary to confirm these conclusions.
\end{abstract}


\section{Introduction}

The well-known correlation between the black hole mass, $M_{\rm BH}$,
and the host galaxy stellar velocity dispersion $\sigma_\star$
(Gebhardt et al. 2000; Ferrarese \& Merritt 2000) implies that black
hole growth and the build-up process of galaxy bulges are closely
related. The fact that the vast majority of AGNs reside in galaxies
with prominent bulges (e.g. Heckman 1980; Ho, Filippenko, \& Sargent
1997; henceforth H97) suggests that bulges play a critical role in the
formation and growth of nuclear black holes.  However, exceptions do
exist, the most notable example being NGC~4395, a nearby Sd galaxy
that lacks a bulge and shows optical and X-ray properties typical of a
Seyfert 1 galaxy (e.g., Filippenko \& Ho 2003; Moran et al. 2005) with
an inferred black hole mass of 3.6 $\times$ 10$^5~{\rm M}_{\odot}$
(Peterson et al.  2005), much less massive than black holes found in
galaxies with massive bulges. Until recently, this galaxy was an
anomaly in that it was the only essentially bulgeless disk galaxy to
harbor a nuclear black hole.  It furthermore contains a nuclear star
cluster (Filippenko \& Ho 2003), providing a unique opportunity to
study the connection between nuclear star clusters and black holes.
More recently, Greene \& Ho (2004; 2007) searched the Sloan
Digital Sky Survey for galaxies with similar intermediate mass black
holes and found more examples of broad-line AGNs, many of which reside
in disk-dominated galaxies that lack a classical bulge. Other more
nearby examples of AGNs or AGN candidates in late-type galaxies have
been found based on infrared, optical, and X-ray studies (Satyapal et
al. 2008; Shields et al. 2008; Ghosh et al. 2008; Barth et al. 2008;
Thornton et al. 2008; Dewangan et al. 2008; Desroches \& Ho 2009).
Although these studies demonstrate that AGNs reside in late-type disk
galaxies, NGC 4395 remains exceptional since it is both near enough
for detailed study and is a truly bulgeless disk galaxy

The investigation of AGNs in bulgeless galaxies is important for the
following reasons. First, it can shed light on the connection between
black holes and galaxy formation and evolution, by addressing the
following outstanding questions:  
How does the black
hole mass and accretion rate relate to the properties of the parent
galaxy in the case of bulgeless galaxies?  Second, since bulgeless
galaxies are thought to be in an early phase of their evolution, they
allow us to probe the mechanisms for growth of supermassive black
holes which are still poorly known.  Constraints on the low-mass end
of the black hole mass function provide the best discriminant among
models that predict the evolution of black holes from extremely high
redshifts to the present time (e.g., Volonteri, Lodato, \& Natarajan
2008).  But before we can carry out any of the above tests, we have to
locate the black holes themselves.

The search for weak nuclear activity associated with black holes in
extremely-late type normal galaxies is observationally challenging
because these galaxies are typically dusty and have significant star
formation rates.  Therefore, such a search is best carried out in
energy bands that do not suffer from heavy dust or gas extinction and
in which contaminating emission from star-forming regions is
relatively weak.  Thus, the mid-IR and hard X-ray bands are well
suited for this purpose and, in fact, studies in these two bands
complement each other well. 

Observations of the mid-IR fine structure lines provide us with one of
the few definitive tools for discovering buried AGNs in dusty
galaxies. AGNs show prominent high excitation fine structure lines
whereas starburst and normal galaxies are characterized by lower
excitation spectra characteristic of \ion{H}{2} regions ionized by
young stars (e.g. Genzel et al. 1998; Sturm et al. 2002; Satyapal et
al. 2004).  One of the key prominent lines is the
[\ion{Ne}{5}]~14$\mu$m line, which cannot be produced in \ion{H}{2}
regions surrounding young stars, the dominant energy source in
starburst galaxies, since even hot massive stars emit very few photons
with energy sufficient for the production of this ion.  The power of
these diagnostics in finding buried AGNs has been convincingly
demonstrated by the discovery of a population of AGNs in galaxies that
display optically ``normal'' nuclear spectra (e.g., Satyapal et
al. 2004).

Once the presence of an AGN is ensured via IR spectroscopy, X-ray
observations represent the most effective means to investigate the AGN
properties, since: 1) The X-rays are produced and reprocessed in the
inner, hottest nuclear regions of the source. 2) The penetrating power
of (hard) X-rays allows them to carry information from the central
engine without being substantially affected by absorption, and to
probe the presence of the putative torus (provided that $N_{\rm H} <
10^{24}~{~\rm cm^{-2}}$). 3) Compared to optical and UV radiation,
X-rays are far less affected by host galaxy contamination.

Using high resolution IRS observations from the \spitzer\ archive,
Satyapal et al. (2007, 2008) started to investigate the IR spectral
properties of late-type (Sc or later) galaxies. Remarkably, of the 33
late-type galaxies with no definitive signatures of AGNs in the
optical, 8 display [\ion{Ne}{5}] emission. This result suggests that
AGNs might be common not only in early-type galaxies but also in
late-type ones, extending the ubiquity of the AGN phenomenon in the
local Universe. In their sample of galaxies showing robust evidence
for a [\ion{Ne}{5}] line, NGC~3621 ($z=0.0024$) is unique in that,
like NGC~4395, it does not show evidence of any bulge component, it
contains a nuclear star cluster (Barth et al. 2009) and is the closest
one, with a distance of 6.6~Mpc bulgeless galaxies based on Cepheid
measurements (Freedman et al. 2001).  Very recently, Barth et
al. (2009), using high quality optical spectra through a small
aperture, detected a Seyfert~2 emission-line spectrum, supporting the
hypothesis that NGC~3621 harbors an accreting supermassive black
hole. Additionally, the measurement of the stellar velocity dispersion
provided an upper limit for the black hole mass of $3\times10^6~{\rm
M_\odot}$.  X-ray observations can strengthen the case for presence of
an accreting black hole in this galaxy and allow us to determine some
of the basic properties of the system, such as the ionizing luminosity
and the accretion rate, given an estimate of the black hole mass.

Here, we report the first X-ray study of \ngc\ based on a 25 ks
\chandra\ observation. The primary goal of this project is to verify
whether this galaxy harbors an accreting supermassive black hole.  As
explained below, our \chandra\ observation reveals the presence of 2
additional bright X-ray sources located $\sim$20\arcsec\ from the
center. Therefore, a secondary goal of this work is to investigate the
nature of these 2 enigmatic X-ray sources.

The outline of the paper is as follows. In $\S~2$ we describe the
observations and data reduction. The X-ray results are reported in
$\S~3$. In $\S~4$ we discuss our findings, and in $\S~5$ we summarize
the main results and draw our conclusions.

\section{Observations and Data Reduction}

\ngc\ was observed with \chandra\ ACIS-S on March 6, 2008.  The
\chandra\ data, provided by the \chandra\ X-Ray Center, were processed
using \verb+CALDB+ v. 3.4.2 following standard criteria.  Only events
with grades 0, 2--4, and 6 and in the energy range 0.5--8 keV were
retained.  We also checked that no flaring background events occurred
during the observations. The extraction of images, light curves, and
spectra, as well as the construction of response matrices, was carried
out following standard procedures.  Background spectra and light
curves were extracted from source-free regions on the same chip of the
source. The counts were extracted from circular regions with radius of
3\arcsec.

The spectral analysis was performed using the {\tt XSPEC v.12.3.0}
software package (Arnaud 1996; Dorman \& Arnaud 2001). For the
brightest source, the spectra were re-binned in order to contain
at least 15 counts per channel, which is appropriate for the use of
the $\chi^2$ statistic. For the second off-nuclear source, given the
low number of counts, we used the Cash statistic (Cash 1979) to fit
the un-binned spectra over an energy range of 0.5--8 keV (in the
observer's reference frame), where the calibration is best known
(Marshall et al. 2005) and the background negligible.  The errors on
spectral parameters are at 90\% confidence level for one interesting
parameter ($\Delta\chi^2,~\Delta C=2.71$).

\section{Results}

\chandra, with its sub-arcsecond spatial resolution, is the best X-ray
satellite to investigate the spatial distribution of the X-ray
emission.  This is crucial for AGN studies and specifically for
detection of weak AGNs, since these are typically associated with
compact un-resolved regions as opposed to starburst activity, which
instead is characterized by extended emission (e.g., Dudik et
al. 2005; Gonz\'alez-Mart\'in et al. 2006; Flohic et al. 2006).

\begin{figure}  
\begin{center}   
 \includegraphics[bb=15 15 550 500,clip=,angle=0,width=10.cm]{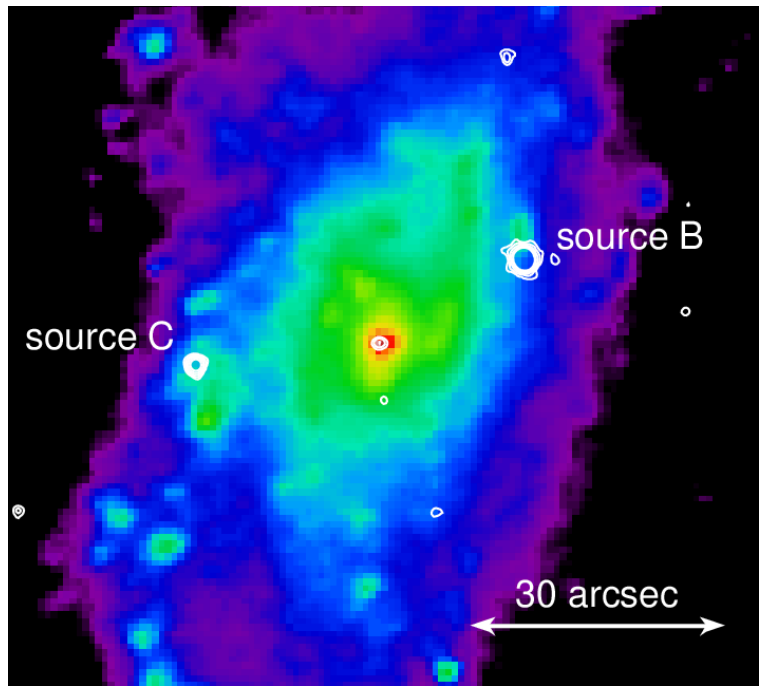} 
\caption{Spitzer IRAC 3.6$\mu$m image of the NGC 3621 galaxy from the SINGS
survey (Kennicutt et al. 2003)
overlaid with contours showing the \chandra\ X-ray point sources.}  
\label{figure:fig1}  
\end{center} 
\end{figure}

\begin{figure}  
\begin{center}   
 \includegraphics[bb=15 15 613 505,clip=,angle=0,width=12.cm]{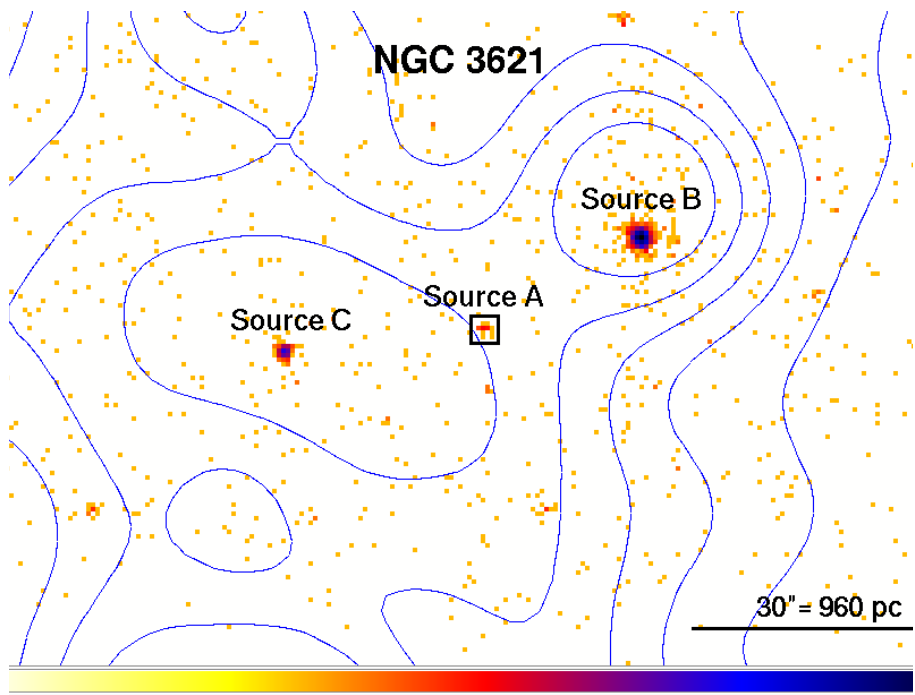}         
\caption{\chandra\ ACIS-S image of \ngc\ in the 0.5--8~keV band with
1.4 GHz contours from a short VLA observation overlaid. The center of
the galaxy, source~A, is shown by the black box. Two off-nuclear
bright sources, source~B and source~C, are located at
$\sim$20\arcsec\ from the center.}
\label{figure:fig2}  
\end{center} 
\end{figure}

Figure~\ref{figure:fig1} shows the \chandra\ contours overlaid with
the \spitzer\ IRAC 3.6$\mu$m 
image of the \ngc\ galaxy from the SINGS survey (Kennicutt et al. 2003).
The nuclear X-ray source shows a clear correspondence with the IR peak.
Figure~\ref{figure:fig2} shows the \chandra\ ACIS-S image of \ngc\
overlaid with 1.4 GHz contours obtained from a VLA\footnote{The
National Radio Astronomy Observatory is operated by Associated
Universities, Inc.\ under a cooperative agreement with the National
Science Foundation.} snapshot observation at 20\arcsec\ resolution
(160 s, archival program AC345). The nuclear X-ray source (hereafter
source~A, marked by a black box) is located at the center of the
image. 

Interestingly, two bright off-nuclear X-ray sources are
detected at $\sim$20\arcsec\ from the center. While no clear IR 
association is evident for the 2 X-ray off-nuclear sources,
the brighter X-ray
source~B may have a radio counterpart in the VLA map with a peak of
5.2 mJy/beam that is $\sim$4\arcsec\ offset (to the north of) the X-ray
source. At $D$=6.6 Mpc, the corresponding 1.4 GHz radio luminosity is
$3 \times 10^{26} {\rm~erg~s^{-1}~Hz^{-1}}$ for an assumed point
source. At the position of the X-ray nucleus and source~C, the radio
map is dominated by diffuse emission from the galaxy and the formal
1$\sigma$ upper limits are 3.9 mJy/beam and 4.9 mJy/beam (the radio
surface brightnesses at the X-ray position), corresponding to point
source limits of $\simlt (2-3) \times 10^{26}
{\rm~erg~s^{-1}~Hz^{-1}}$. A higher resolution VLA 8.5 GHz snapshot
(107 s, archival program AR555) did not reveal any significant
emission at these locations with point source limits of 0.5 mJy
(5$\sigma$) at 0.53\arcsec\ $\times$ 0.22\arcsec\ resolution. The
vastly differing resolutions of the two radio maps preclude any
practical constraints on the radio source spectra.

At first glance the symmetric placement of the two off-nuclear sources
relative to respect to the center of the galaxy, and their apparent
association with radio emission, suggests that they may be related to
the central source. They resemble hot-spots produced by the
interaction of bipolar outflows with the interstellar medium. However,
the results of our analysis of their time-averaged X-ray spectra (see
below) seem to rule out this scenario.

\subsection{Source A: The Central Source}

The X-ray 
central region is located at RA=11$^{\rm h}$18$^{\rm
m}$16.51$^{\rm s}$, DEC=-32\arcdeg\ 48\arcmin\ 50.4\arcsec\
(the coordinates here and in the rest of the paper refer to the J2000
epoch, and have been determined using the \verb+celldetect+ task in CIAO). 
Note that the \chandra\ position of the central source is fully
consistent with the \ngc\ coordinates  from
the {\it 2MASS} catalog, RA=11$^{\rm h}$18$^{\rm
m}$16.50$^{\rm s}$, DEC=-32\arcdeg\ 48\arcmin\ 50.6\arcsec\
(Skrutskie et al. 2006).

\begin{figure}  
\begin{center}   
 \includegraphics[bb=15 15 470 345,clip=,angle=0,width=10.cm]{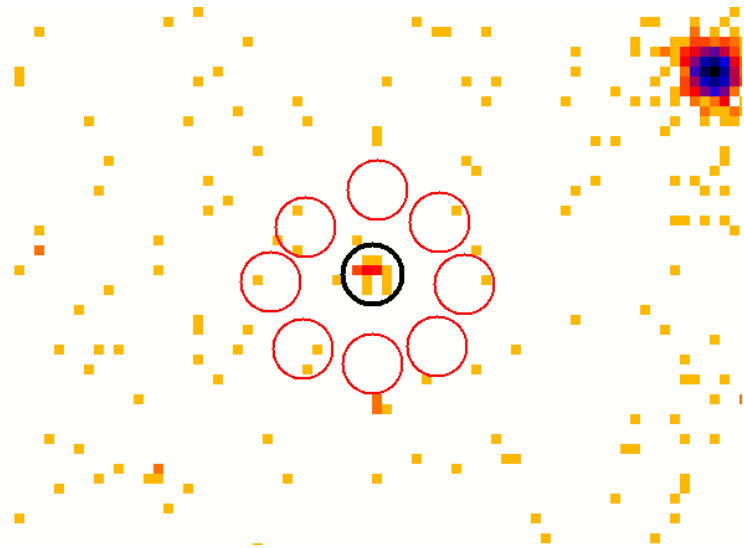} 
\caption{\chandra\ ACIS-S  image of \ngc. The extraction region for the 
central source is indicated by the black thick circle. The surrounding 
circles are the background regions.}  
\label{figure:fig3}  
\end{center} 
\end{figure}

Using the
\verb+CIAO+ tool \verb+dmextract+, we extracted the X-ray counts from
a central circular region of radius of 1.5\arcsec\ (black thick circle
in Figure~\ref{figure:fig3}) and from 8 background regions surrounding
it, as shown in Figure~\ref{figure:fig3}.  After background
subtraction, we detect $21\pm6$ X-ray counts from the source, which is
significant at the 3.7 $\sigma$ level.  The uncertainty was calculated
using the approach of Gehrels (1986), which is appropriate for a small
number of counts.

Unfortunately, with such few counts, it is impossible to spectrally
characterize source~A; neither direct spectral fitting nor hardness
ratio calculations yield useful information. Examining the morphology
in several, narrow energy bands only reveals that most of the counts
($\sim$70\%) are in the energy range 0.8--1.8 keV, which coincides
with the peak of the \chandra\ ACIS effective area curve.  Therefore,
we resort to deriving some estimates of the absorbed flux and
corresponding intrinsic luminosity using the simulation tool
\verb+PIMMS+.

Assuming a power-law model with photon index $\Gamma$ ranging from 1.7
to 2.0 (which is typical for AGNs) and a Galactic column density of
$N_{\rm H}=5.74\times10^{20}$\nh, we obtain an absorbed flux of
$F_{\rm 0.5-8\, keV}\sim 6.5\times 10^{-15}$ \flux\ ($F_{\rm 2-10\,
keV}\sim 4\times 10^{-15}$ \flux). The Seyfert 2 spectroscopic
classification and the relatively large bolometric luminosity,
inferred by both IR and optical studies (Satyapal et al. 2007; Barth
et al. 2009), suggest the presence additional absorption, local to
\ngc. In order to infer plausible values for the intrinsic luminosity,
we assumed some additional intrinsic absorption with column densities
ranging from $5\times10^{21}$ to $5\times10^{23}$~\nh. The lower value
of $N_{\rm H}$ is consistent with the one derived from spectral
fitting of the 2 off-nuclear sources (see below), whereas the upper
value is 
a typical value found in non Compton-thick Seyfert 2 galaxies
(see, e.g., Figure 3 of Bassani et al. 1999).
 The resulting values of the estimated
luminosities, which mostly depend on the choice of intrinsic $N_{\rm
H}$ (variations of $\Gamma$ cause luminosity differences of the order
of a few percent only) are $L_{\rm 0.5-8\, keV}\sim 7\times 10^{37}$
\lum\ ($L_{\rm 2-10\, keV}\sim 5\times 10^{37}$ \lum) and $L_{\rm
0.5-8\, keV}\sim 3\times 10^{39}$ \lum\ ($L_{\rm 2-10\,keV}\sim
2\times 10^{39}$ \lum), for the moderate and heavy absorption
scenarios, respectively.

In summary, the \chandra\ observations reveal the presence of a weak
point-like X-ray source at the center of \ngc, which confirms the
presence of an AGN. However, the paucity of counts hampers the study
of the nature of this source and more specifically whether this system
is heavily absorbed or not.

\subsection{Source B: the Brightest Off-Nuclear Source} 

Source~B is fairly X-ray bright (more than 2000 counts were collected
during the 20.6 ks net exposure) located at RA=11$^{\rm h}$18$^{\rm
m}$15.16$^{\rm s}$, DEC=-32\arcdeg\ 48\arcmin\
40.6\arcsec. Since Source~B is much brighter than the other sources in the
\chandra\ field of view, one may have the impression that its
emission is spatially extended. However, a comparison of the brightness
radial profile with the ACIS point spread function reveals that X-ray spatial
distribution is fully consistent with that of a point-like source.

\begin{figure}  
\begin{center} 
\includegraphics[bb=30 280 580 510,clip=,angle=0,width=14.cm]{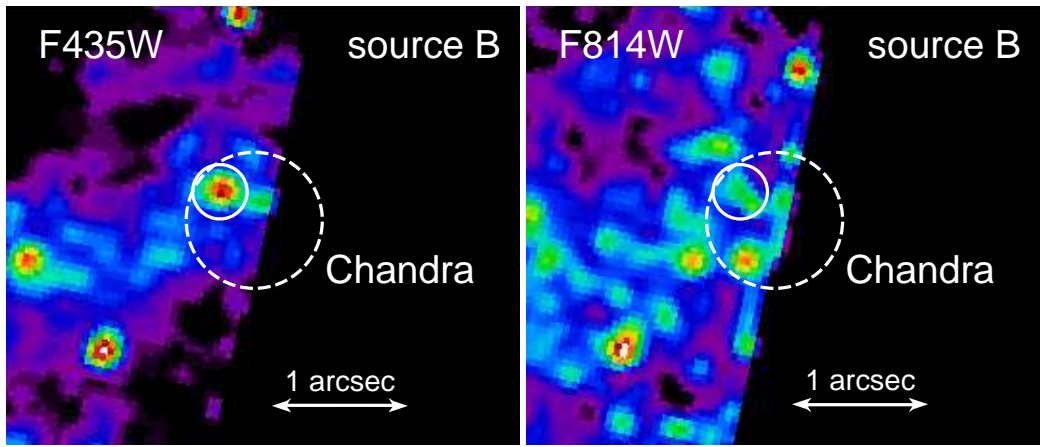}
\caption{\hst\ ACS WFC images (in 2 of the 3 observed bands) centered 
around the position of
Source B, the brightest X-ray off-nuclear that is  
indicated with r=0.5" dashed circles. The images were
smoothed with a Gaussian with $\sigma$ =
2 pixels (0.098"). The edge of the WFC field of view is obvious in all
images with only about $\sim$1/2
of the Chandra source B circle covered. Detected optical sources
tabulated in Table~1 are indicated
with solid r=0.2" circles.}
\label{figure:fig4}
\end{center}
\end{figure}

Figure~\ref{figure:fig4} shows the  Hubble Space Telescope
(\hst) ACS WFC images 
centered around the position of Source B. The position of the
off-nuclear X-ray source was observed at
the edge of the
field of \hst\ images obtained in the F435W, F555W (not shown), 
and F814W filters (each with 3x360s exposures; program 9492)
on 2003 February 16. 
Aligning the X-ray and optical nucleus in the \chandra\ and \hst\ images,
we find a faint
optical source 0.33" from the center of the X-ray position of source
B, although only
the eastern $\sim$1/2 of the X-ray error circle was covered in the WFC images.
Ulitizing r=0.2" circles, we measured count rates for the optical
source including
four surrounding source-free regions for background estimation. The
optical source is
a 4.5$\sigma$ detection in the F435W image, a lower significance detection
(2.8$\sigma$) in the F555W image, and is undetected in the F814W image, where 1
$\sigma$ errors were determined by calculating the standard deviation of the
background count rates. Count rates were converted to flux density utilizing 
the
PHOTFLAM and PHOTPLAM keywords in the FITS header and the aperture corrections
tabulated in Sirianni et al. (2005), and are reported in Table 1.

\begin{figure}  
\begin{center} 
\includegraphics[bb=64 24 545 704,clip=,angle=-90,width=9.cm]{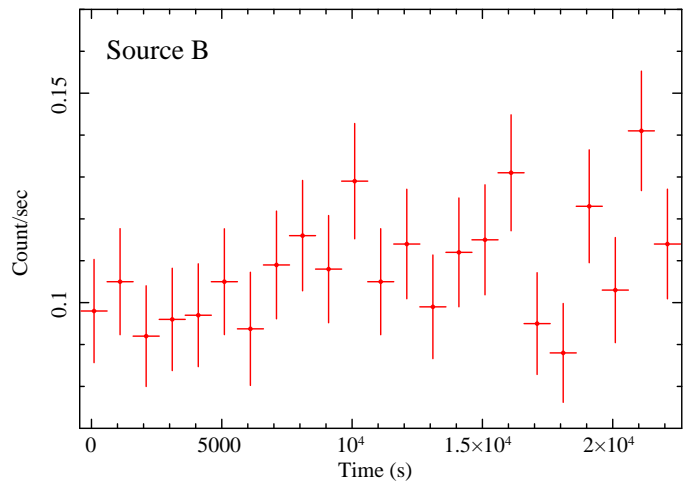}
\caption{\chandra\ ACIS-S light curve of source~B in the 0.5--8~keV energy 
band; time bins are 1000 s.}
\label{figure:fig5}
\end{center}
\end{figure}

The relatively high X-ray count rate of Source B allows us to 
investigate not only the
X-ray spectral properties via direct fitting of the spectrum but also
the temporal properties.  We have extracted the light curve in the
0.5--8 keV energy range, from a circular region with a radius of
3\arcsec\ centered on source~B.  This light curve is shown in
Figure~\ref{figure:fig5}. Although there appear to be some
low-amplitude fluctuations toward the end of the observation,
according to a $\chi^2$ test the light curve is consistent with the
hypothesis that the source flux is constant during the \chandra\
pointing ($\chi^2$/dof=24.20/22, $P_{\chi^2}$= 0.34).

\begin{figure}
\begin{center}
\includegraphics[bb=40 2 567 704,clip=,angle=-90,width=9.cm]{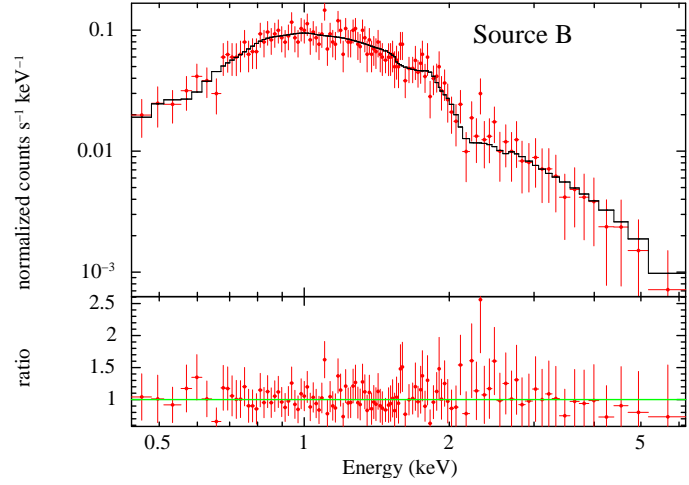}
\caption{Spectrum of source~B and data/model ratio to a simple power-law 
model modified by intrinsic photoelectric absorption.}
\label{figure:fig6}
\end{center}
\end{figure}

The X-ray spectrum of source~B, obtained from the same extraction
region as the light curve, is poorly fitted by a simple power law with
Galactic absorption ($\chi^2$/dof=154.12/112), showing a deficit at
low energies indicative of additional absorption. Indeed, by leaving
the column density of the absorber free to vary, a much better fit is
obtained ($\chi^2$/dof=50.47/111) with the following parameters:
$N_{\rm H}= (3.4\pm0.6)\times10^{21}$ \nh\ and $\Gamma=2.7\pm0.3$. The
spectrum of source~B, with the best fit overlaid, and the
data-to-model ratio are shown in Figure~\ref{figure:fig6}.  The
corresponding absorbed flux is $F_{\rm 0.5-8\, keV}=7\times10^{-13}$
\flux\ and the intrinsic luminosity, assuming that source~B is located
at the same distance as \ngc, $L_{\rm 0.5-8\, keV}=7\times10^{39}$
\lum\ ($L_{\rm 2-10\, keV}=2\times10^{39}$ \lum).
We have also tried to add a thermal component (described by the XSPEC
model \verb+apec+). Although the overall fit remains acceptable
($\chi^2$/dof=48.97/108), the spectral parameters of the thermal
component ($kT$ and abundances) remain largely unconstrained.  

The large luminosity, well in excess to the Eddington limit for a neutron
star, suggests that source~B can be a black hole system. As a
consequence, the X-ray emission is likely to be produced by the
Comptonization of seed photons emitted by the underlying accretion disk. It is
therefore instructive  (and indeed a necessary step to constrain 
$M_{\rm BH}$ in this system with the method described in $\S4.2$) to fit 
the \chandra\ spectrum of source~B with the
physically motivated \verb+BMC+ model instead of a phenomenological
power law (see Titarchuk et al. 1997 for further details). 

The Bulk Motion Comptonization model (BMC) is a generic Comptonization 
model able to describe equally well bulk motion and thermal Comptonization,
although it was historically developed to describe the Comptonization of 
thermal seed photons by a relativistic converging flow (Titarchuk et al. 1997).
The BMC model is characterized by 4 free parameters: the temperature of the
thermal seed photons $kT$, the energy spectral index $\alpha$, a parameter 
$\log(A)$ related to the Comptonization fraction $f$  by the relation 
$f=A/(1+A)$ (where $f$ is the ratio of Compton scattered photons over 
the seed photons), and the normalization $N_{\rm BMC}$.
In simple words, the BMC model convolves the thermal seed photons and a
{\it generic} Comptonization  Green's function producing
a power law. As a consequence, this model adequately fits 
X-ray spectra of accreting black holes, which at the zeroth order are always
characterized by a hard power law, which is widely believed to be
produced by the Comptonization of a seed thermal component.
In addition to being a comprehensive Comptonization model,
the BMC model presents 2 important advantages with respect to the power law
model (PL): 1) Unlike the PL, which is a phenomenological model, the BMC 
parameters are computed in a self-consistent way; 2) Unlike the PL, the power 
law produced by BMC does not extend to arbitrarily low energies and thus does 
not affect the normalization of the thermal component nor the amount of local 
absorption, which is often present around accreting objects.

The resulting best fit ($\chi^2$/dof=49.90/110), which is statistically
indistinguishable from the PL fit, yields the following parameters:  
$N_{\rm H}= (3.1\pm0.1)\times10^{21}$ \nh, $kT=0.1\pm0.1$ keV (the
temperature of the seed photons), $\alpha=1.6\pm0.3$ (the spectral
index, which is related to the photon index by $\Gamma=\alpha+1$),
$\log A=2$ (fixed at the best fit value), and 
$N_{\rm BMC}=1.2_{-0.4}^{+6.8}\times 10^{-5}$, which is
in units of $(L/10^{39}\; {\rm erg~s^{-1}})(10\;{\rm kpc}/d)^2$ with
$L$ and $d$ being the luminosity and distance of the object,
respectively.

\subsection{Source C: the second off-nuclear source} 
\begin{figure}  
\begin{center} 
\includegraphics[bb=30 280 580 510,clip=,angle=0,width=14.cm]{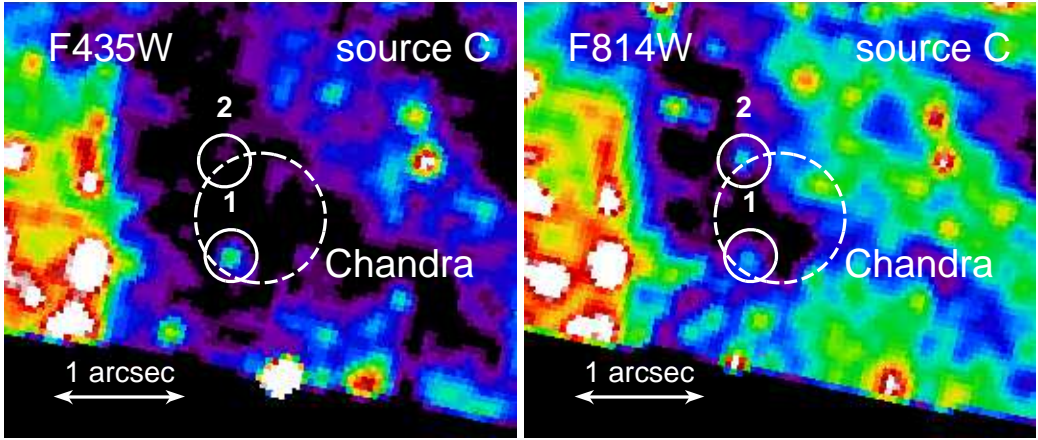}
\caption{\hst\ ACS WFC images (in 2 of the 3 observed bands) centered 
around the position of
Source C, indicated with r=0.5" dashed circles. The images were
smoothed with a Gaussian with $\sigma$ =
2 pixels (0.098").  Detected optical sources
tabulated in Table~1 are indicated
with solid r=0.2" circles.}
\label{figure:fig7}
\end{center}
\end{figure}

The measured X-ray position of 
Source~C is RA=11$^{\rm h}$18$^{\rm m}$18.23$^{\rm s}$,
DEC=-32\arcdeg\ 48\arcmin\ 53.0\arcsec. Figure~\ref{figure:fig7}
shows the \hst\ ACS WFC images (described in the previous subsection)
centered around the position of Source C.
A faint optical source is detected 0.37" away from
the \chandra\ position
in all 3 filters. An additional redder source is visible 0.5" away but
detected only in
the F814W filter image. Additional ACS images from the HST same
program obtained on Feb 3  confirm the two optical
sources. The
photometric results from this latter dataset are within the 1$\sigma$
errors with the
exception of the F435W measurement of C1 where we measure a flux that
is 2$\sigma$ lower
than that reported in Table~1.

The moderately low X-ray count rate ($\sim$200 counts collected during the net
20.6~ks exposure) allows a direct fitting of the un-binned spectrum
using the $C-$statistics, which can be used to estimate parameters 
values and confidence regions but does not  provide a goodness-of-fit.
To this aim, on can use the \verb+XSPEC+ command \verb+goodness+ that
performs Monte Carlo simulations of spectra based on the chosen model.
This procedure yields the percentage of simulations with the fit statistic 
lower than that for the original data. If the best-fitting model is a 
good representation of the data the percentage  should be around 50\%;
values $\ll$50\% indicate that the data are over-parameterized, whereas 
values close to 100\% indicate that the fit is poor.

\begin{figure}
\begin{center}
\includegraphics[bb=40 2 567 704,clip=,angle=-90,width=9.cm]{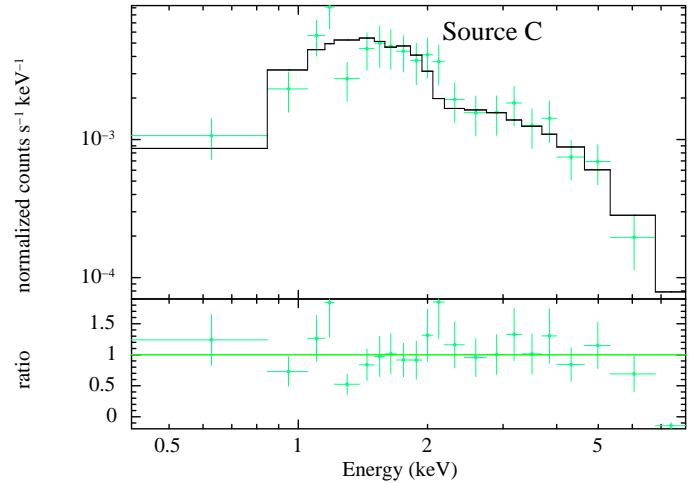}
\caption{Spectrum of source~C and data/model ratio to a simple power-law 
model modified by intrinsic photoelectric absorption.}
\label{figure:fig8}
\end{center}
\end{figure}

Source C appears to be reasonably well fitted with a
simple power-law model with absorption in excess to the Galactic
$N_{\rm H}$ (see Figure~\ref{figure:fig8}), although the 
\verb+goodness+ command yields 74\% indicating that this model
is only marginally acceptable.
The best fit parameters
are: $N_{\rm H}= (5.1_{-1.9}^{+2.5})\times10^{21}$ \nh\ and
$\Gamma=1.6\pm0.4$.  The corresponding absorbed flux is $F_{\rm
0.5-8\, keV}=1.1\times10^{-13}$ \flux\ and the intrinsic luminosity,
assuming that source~C is located at the same distance as \ngc,
$L_{\rm 0.5-8\, keV}=7\times10^{38}$ \lum\ ($L_{\rm 2-10\,
keV}=6\times10^{38}$ \lum).  

Also in this case we have tried the
Comptonization \verb+BMC+ model instead of a phenomenological power
law. The resulting best fit parameters are: $N_{\rm H}=
(5.0\pm2.2)\times10^{21}$ \nh, $kT=0.06_{-0.05}^{+0.08}$ keV,
$\alpha=0.5\pm0.2$ , $\log A=-1.15$, $N_{\rm BMC}=(6.3\pm5.0)\times
10^{-6}$.  The \verb+goodness+ command (98\%) indicates that 
this model is statistically worse than the power law. However, as
explained above, the results from the fitting of the \verb+BMC+ model
represent the first necessary step to attempt an estimate of $M_{\rm BH}$
in this putative black hole system.

\section{Discussion}

\subsection{The central source}

The primary goal of our X-ray investigation is to verify whether \ngc\
hosts a buried AGN, as suggested by recent IR and optical
spectroscopic studies.  The presence of a weak X-ray source coinciding
with the center of the galaxy seems to be consistent with this
scenario. Unfortunately, however, the very low count rate severely
hampers our investigation, preventing any X-ray spectral
characterization of this source. Nevertheless, combining the \chandra\
results with findings from recent IR and optical spectroscopic
studies, we may address several important questions.

Before proceeding further, it is worth summarizing the main results
from longer wavelength studies that will be used in our
discussion. From the IR study, we will use the bolometric luminosity,
$L_{\rm bol}=5\times10^{41}$ \lum, which was derived from the $L_{\rm
[Ne\,V]}-L_{\rm bol}$ correlation (Satyapal et al. 2007). From
the optical investigation of Barth et al. (2009) we will make use of
the following information: 1) \ngc\ is spectrally classified as a
Seyfert 2; 2) The measured [\ion{O}{3}]~$\lambda5007$ flux is $F_{\rm
[O\,III]}=(22.4\pm0.4)\times 10^{-16}$ \flux, which, based on inspection of the
two-dimensional optical line ratios, probably encompasses only 10\% of
the narrow line region (NRL). Note that if the last hypothesis is
correct, the bolometric luminosity derived by $L_{\rm [O\, III]}-L_{\rm
bol}$ correlation is fully consistent with the value inferred from
$L_{\rm [Ne\,V]}$; if not, $L_{\rm bol}$ derived from $L_{\rm [O\, III]}$
is lower by a factor of $\sim$10.

The first outstanding question is whether the central source is
heavily absorbed or intrinsically weak. Based on the \chandra\ count
rate, the 2--10 keV luminosity we have inferred has a range of over 2
orders of magnitude (from a few $\times 10^{37}$ to a few $\times
10^{39}$ \lum), depending on the adopted intrinsic $N_{\rm H}$
($5\times10^{21}$ and $5\times10^{23}$\nh, respectively). 

We can estimate the $L_{2-10~\rm keV}$ luminosity using the $L_{[Ne\,V]}$ 
luminosity assuming the $L_{[Ne\,V]}/L_{2-10~\rm keV}$ ratio in standard AGN.  Using the  limited number of standard AGN with both [NeV] and 2-10 keV 
observations (based on  recent Spitzer observations of AGN from 
Dudik et al. 2007 and Gorjian et al. 2008), we find an average ratio of 
$L_{[Ne\,V]}/L_{2-10~\rm keV}\simeq 10^{-3}$. This result, combined with
the estimated value of $L_{[Ne\,V]}$ for \ngc, $\sim5\times10^{37}$ \lum, 
predicts a 2--10 keV luminosity of the order of $10^{40}$ \lum, which is
more than two orders magnitude greater than the observed luminosity in the
weakly absorbed scenario.  Therefore, this result strongly suggests that 
NGC 3621 is heavily absorbed.

A second indirect
way to test the value of $L_{\rm X}$ employs the correlation between
$L_{\rm X}-L_{\rm [O\, III]}$.  Using the results from Mulchaey et
al. (1994) for a sample of Seyfert galaxies, and, more specifically,
utilizing the best-fit linear correlation derived by their Figure 3c,
we obtain a luminosity of $L_{\rm X}=9.8\times10^{39}$ \lum, which is
in broad agreement with the heavily absorbed scenario. 

A third alternative
way to constrain $N_{\rm H}$ (and hence $L_{\rm X}$) is based on the
$F_{\rm X}/F_{\rm [O\, III]} - N_{\rm H}$ diagram proposed by
Guainazzi et al.  (2005) and derived from the work of Bassani et
al. (1999) on Seyfert 2 galaxies.  In this diagram, which shows a
clear anti-correlation between $F_{\rm X}/F_{\rm [O\, III]}$ and
$N_{\rm H}$, $F_{\rm X}$ is the {\it absorbed} 2--10 keV flux whereas
$F_{\rm [O\, III]}$ is {\it corrected} for absorption based on the
$H\alpha/H\beta$ ratio (see the appendix of Bassani et al. 1999 for
details).  Depending on whether $F_{\rm [O\, III]}$ encompasses 10\%
or the entire [\ion{O}{3}] luminosity of the NLR, the $F_{\rm
X}/F_{\rm [O\, III]}$ ratio lies between 0.1 and 1, which in turn
translates into $N_{\rm H}$ ranging between $\sim 10^{23}$ and $\sim
10^{24}$~\nh.  This result again lends support to the heavily absorbed
scenario. Finally, this scenario is supported by the direct comparison
between $L_{\rm X}$ and $L_{\rm bol}$. Indeed, if we use the value
obtained with moderate absorption, $5\times10^{37}$\lum, we derive a
bolometric correction factor of $\kappa_{\rm 2-10\, keV}\equiv L_{\rm
bol}/L_{\rm 2-10\,keV} > 1000$, which is unreasonably high and
inconsistent with any reasonable spectral energy distribution (SED)
for AGNs.

Adopting the heavily obscured scenario, we can now try to constrain
the most important parameters for a black hole system, namely the
accretion rate, $\dot m$, and the black hole mass, $M_{\rm
BH}$. Assuming that \ngc\ has an X-ray luminosity of few $\times
10^{39}$ \lum, the bolometric correction factor becomes $\kappa_{\rm
2-10\, keV}\ga 100$.  This result directly translates into an estimate
the Eddington ratio value (and hence on $\dot m$) by virtue of the
existing positive correlation between the bolometric correction factor
$\kappa_{\rm 2-10\, keV}$ and $L_{\rm bol}/L_{\rm Edd}$ (Vasudevan and
Fabian 2009). According to Vasudevan and Fabian (2009), this
bolometric correction is characteristic of systems accreting at higher
rate: $L_{\rm bol}/L_{\rm Edd} \ga$ 0.2. This finding in turn can be
used to constrain the black hole mass: Assuming that $L_{\rm
bol}=5\times10^{41}$ \lum and $L_{\rm bol}/L_{\rm Edd}\simeq 0.2$, we
obtain $M_{\rm BH}\simeq 2\times 10^4 {\rm~ M_\odot}$, which is a
fairly low value and apparently below the extrapolation of the $M_{\rm
BH}-\sigma$ correlation (see, e.g., Fig. 10 of Barth et al. 2009). The
small $M_{\rm BH}$ coupled with the high $\dot m$ appears to be in
line with the scenario where bulgeless galaxies host small
supermassive black holes in their early stage that undergo a phase of
very vigorous accretion, and is consistent with recent findings from
Greene \& Ho (2007) and Barth et al. (2008) on Seyfert 1 and Seyfert 2
galaxies, respectively.

We should point out that the scenario of a heavily absorbed AGN in NGC
3621 is not without caveats.  The correlation between the
[NeV] luminosity and the bolometric luminosity in standard AGN
shows a scatter 0.44 dex, which corresponds to a factor of $\sim$3 
uncertainty in $L_{\rm bol}$ derived by $L_{\rm [NeV]}$ 
(Satyapal et al. 2007).  Furthermore, its
applicability at this luminosity range is uncertain.  In addition the
fraction of the [OIII] luminosity encompassed by the
optical spectrum obtained by Barth et al. (2009) and attributable to
the AGN is also uncertain.  However, taken collectively, there is
strong support for the hypothesis that NGC 3621 harbors a buried AGN
with luminosity of the order of a few times $10^{41}$ \lum.

\subsection{The off-nuclear sources}

The secondary goal of this work is to investigate the nature of the 2
bright off-nuclear sources detected by \chandra. 
Based on the X-ray luminosity values derived from the
spectral analysis ($L_{\rm 2-10\,keV}= 2\times10^{39} ~{\rm and}~
6\times10^{38}$ \lum\, for sources B and C, respectively), with the
assumption that these sources are located within \ngc\ (i.e., at a
distance of 6.6 Mpc), we can already derive the first important
conclusion: these are ultraluminous X-ray sources and their compact
objects are likely black holes, since their X-ray luminosity exceeds
the Eddington limit for neutron stars.

The lack  of strong optical/UV counterparts (see Table 1), 
the fact that they have similar intrinsic
absorption that is well in excess with respect to the Galactic
value, and the apparent association with the \ngc\ radio emission
suggest that these sources are not located at
significantly smaller distances than \ngc.

However, a priori, we cannot rule out that they are
background AGN that appear to be located close to the center of
\ngc\ just by chance. To test this hypothesis, we have performed the
following test. Utilizing the results of the \chandra\ spectral analysis 
we have derived the monochromatic X-ray luminosity 
$l_{\rm 2~keV}$ for different values of redshift ranging between $z=0.05-1$.
From the well-known tight correlation $l_{\rm 2~keV}-l_{\rm 2500\AA}$
for standard AGN (we used equation 1c from Steffen et al. 2006)
we have determined the corresponding values of $l_{\rm 2500\AA}$ and the
relative fluxes. Finally, the expected values for the flux at the wavelengths
probed by the \hst\ observations
have been obtained by converting the 2500\AA\ fluxes assuming a typical 
slope of 0.7 ($f_\nu\propto\nu^{-0.7})$. 

For Source B the predicted optical 
flux ranges between 230--370 $\mu$Jy (depending on the optical band) at 
$z=0.05$ and 6-9 mJy at $z=1$. For the weaker X-ray Source C, we obtain 
optical 
fluxes of the order 14-21 $\mu$Jy at $z=0.05$ and 345--535 $\mu$Jy at $z=1$.
Since these expected values for standard AGN are at least 2 orders of 
magnitude larger than the measured optical fluxes reported in Table 1, we
can rule out the hypothesis of background AGN.
We also applied extinction corrections by assuming the
absorption column density derived from the X-ray spectral fits, $N_{\rm H}$,
and converting it into optical extinction with the relation $A(V)/N_{\rm H}=
5.3\times 10^{-22}$ (Cox \& Allen 2000). This corresponds to 
$A(V)=$1.8 and 2.7, which in turn translates into flux correction factors of 
$\sim$5 and $\sim$10 for source B and C, respectively. Nevertheless, the
measured optical
fluxes are still inconsistent with the predicted values for standard AGN.

We can therefore reasonably assume that both off-nuclear sources reside
in \ngc. At a distance of 6.6 Mpc, the measured fluxes of their
putative optical counterparts correspond to optical luminosities of the
order of $\sim10^{36}-10^{37}$ \lum, which, combined with the estimated X-ray
luminosities, appear to be in line with typical values of
$L_{\rm X}/L_{\rm opt}$
measured in X-ray binaries (e.g., Ritter \& Kolb 2003).  
In the hypothesis that 
these sources
are indeed hosted by \ngc, it is important to determine whether they are
analogs of Galactic microquasars like GRS~1915+105, or intermediate
black holes (IMBHs) close to the center of the galaxy.

In order to shed light on the nature of these sources, and
specifically derive useful constraints on their $M_{\rm BH}$ and
distance, we will make use of two different methods, that are based on
the X-ray spectral fitting results with the \verb+BMC+ model. Both
methods rely on the generally accepted assumption that the X-ray
spectrum is produced by Comptonization of a thermal component
associated with an accretion disk.  The first method assumes a
specific model for the accretion disk based on the seminal work from
Shakura \& Sunyaev (1973) and interpret $kT$ as the color temperature
of the accretion disk (which is related to the effective disk
temperature by a ``hardening factor'' $T_h$). On the other hand, the
second method only assumes that the underlying physics in black hole
systems is the same irrespective of the scale.  Therefore, with the
second method $M_{\rm BH}$ can be directly derived by scaling
self-similar relationships between spectral (and temporal) parameters.
For further details, see Shaposhnikov \& Titarchuk (2009).

The first method has been introduced by Shrader \& Titarchuk (1999)
and successfully applied to derive $M_{\rm BH}$ in several Galactic
black holes (GBHs), in intermediate BHs, and also in a sample of
Narrow Line Seyfert 1 galaxies (see Shrader \& Titarchuk 2003 for
details). The second method, on the other hand, is based on the
extension of the scaling technique presented by Titarchuk \& Fiorito
(2004) and Shaposhnikov \& Titarchuk (2007).  Specifically, it is
based on the universal scalable relationship between the photon index
and the normalization of the \verb+BMC+ model, which appears to be
characteristic of all GBHs during their spectral transitions
(Shaposhnikov \& Titarchuk 2009). 

In simple terms, the second method
can be summarized by the following steps. 1) Construct a $\Gamma -
N_{\rm BMC}$ plot for a GBH of known mass and distance, which will be
used as reference (hereafter denoted by the subscript {\it r}). 2)
Compute the normalization ratio between the target of interest
(hereafter denoted by the subscript {\it t}) and the reference object
$N_{\rm BMC,t}/N_{\rm BMC,r}$ at the corresponding value of
$\Gamma$. 3) Derive the black hole mass using the following equation
\begin{equation}
M_{\rm BH,t}=M_{\rm BH,r}\times 
(N_{\rm BMC,t}/N_{\rm BMC,r})\times (d_t/d_r)^2 \times f_G
\end{equation}
where $M_{\rm BH,r}$ is the black hole mass of the reference object,
$N_{\rm BMC,t}$ and $N_{\rm BMC,r}$ are the respective normalizations
for target and reference objects, $d_t$ and $d_r$ are the
corresponding distances, and $f_G=\cos\theta_r/\cos\theta_t$ is a
geometrical factor that depends on the respective inclination
angles. The above formula is readily obtained by considering that a)
the normalization is a function of luminosity and distance: $N_{\rm
BMC}\propto L/d^2$; b) the luminosity can be expressed by $L\propto
\eta M_{\rm BH} \dot m$, where $\eta$ is the radiative efficiency, and
c) assuming that different sources in the same spectral state (defined
by the photon index) have the same $\eta$ and $\dot m$.

\begin{figure}
\begin{center}
\includegraphics[bb=40 30 355 300,clip=,angle=0,width=9.cm]{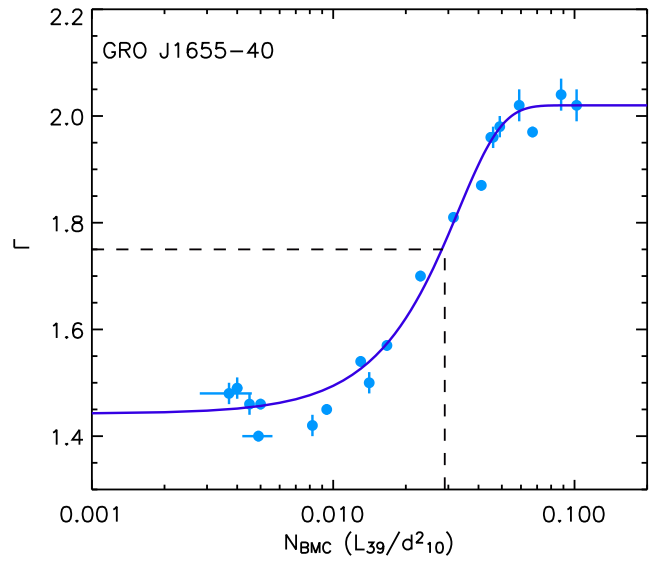}
\caption{Photon index plotted versus the normalization of the BMC
model for the microquasar GRO~J1655-40 during the decay of the 2005
outburst (adapted from Shaposhnikov \& Titarchuk (2008). The thick
continuous line represents the best fitting function of the spectral
trend.  The dashed lines illustrate how we determine the value of
$N_{\rm BMC,r}$ that we use in equation~(1) to estimate the black hole
mass of the BLRG of 3C~390.3.}
\label{figure:fig9}
\end{center}
\end{figure}

To illustrate this method, in Figure~\ref{figure:fig9} we show the
$\Gamma - N_{\rm BMC}$ diagram for GRO J1655-40, which is the primary
reference source since the parameters of this system are the most
tightly constrained: $M_{\rm BH}/M_{\odot}=6.3\pm0.3$, $i=70^o\pm1^o$,
$d=3.2\pm0.2$ kpc (Greene et al. 2001; Hjellming \& Rupen 1995; but
see also Foellmi et al. 2006 for a discording view on the distance of
GRO~J1655-40).

Since, unlike the first method from Shrader \& Titarchuk, this method
is relatively new and has never been tested before for AGNs or IMBHs,
we apply it first to an AGN of known mass, as a check. We chose the
broad-line radio galaxy (BLRG) 3C 390.3 (z=0.056 corresponding to a
luminosity distance of 247 Mpc), because, it has a well constrained
mass as well as a well constrained disk inclination.  One black hole
mass estimate of $M_{\rm BH}=(5\pm1)\times 10^8~{\rm M_\odot}$ is
based on the velocity dispersion (Nelson et al. 2004; Lewis \&
Eracleous 2006) and another estimate of $M_{\rm BH}=(2.9\pm0.6)\times
10^8~ {\rm M_\odot}$ based on reverberation mapping (Peterson et
al. 2004)-- An inclination angle of 25\arcdeg$\,<i<\,$35\arcdeg\ is
derived from the radio data (Eracleous, Halpern \& Livio 1996,
Giovannini et al. 2001, Lewis et al. 2005), while an angle of
$26^\circ {}^{+4^\circ}_{-2^\circ}$ is derived from the profiles of
the double-peaked optical emission lines (Eracleous \& Halpern 1994).
Fitting the \xmm\ EPIC pn spectrum of 3C~390.3 with the \verb+BMC+
model (and a Gaussian line to account for the \feka\ emission), we
obtain: $kT=0.106\pm0.003$ keV, $\alpha=0.75\pm0.01$ ,
$\log(A)=0.66\pm0.03$, and $N_{\rm BMC}=(8.7\pm0.2)\times
10^{-5}$. From Figure~\ref{figure:fig6}, we infer that at
$\Gamma=1.75$ the reference normalization is $N_{\rm BMC,r}\simeq
0.3$.  Inserting this result and the corresponding quantities in
equation~(1) we obtain $M_{\rm BH}\simeq 1.5\times 10^8~{\rm
M_\odot}$. Using $d_{\rm GROJ1655-40}= 1.7$ kpc, as suggested by
Foellmi et al. (2006), the resulting mass increases by a factor
$\sim$3.5, yielding $M_{\rm BH}\simeq 5\times 10^8~{\rm M_\odot}$,
which is in good agreement with the accepted value.

Since the value of $M_{\rm BH}$ derived for 3C~390.3 is reasonably
close to the estimates obtained from optical data, we can try to apply
this procedure (hereafter method~2 for simplicity) and the Shrader \&
Titarchuk method (method~1) to the 2 X-ray enigmatic sources detected
by \chandra.  Using the results from the spectral fitting with the
\verb+BMC+ model, for source~B we obtain $M_{\rm BH}\simeq
(1.4\times10^3 - 1.5 \times 10^4)~{\rm M_\odot}$ and $M_{\rm BH}\simeq
(3\times10^2 - 3.4 \times 10^3)~{\rm M_\odot}$, using method~1 and
method~2, respectively. The ranges of $M_{\rm BH}$ reflect the
relatively large uncertainties of the spectral parameters. Similarly,
for source~C, the two methods yield respectively $M_{\rm BH}\simeq
(1.8\times10^3 - 1.3 \times 10^4)~{\rm M_\odot}$ and $M_{\rm BH}\simeq
(2.3\times10^3 - 2 \times 10^4)~{\rm M_\odot}$.

These findings are important for a number of reasons. First, the
consistency between the BH mass estimates obtained with 2 different
methods lends support to the validity of the 2 methods. Second, since
the methods are independent, the mass obtained from one method can be
used in the second method to verify the distance of the object, and
the similarity between $M_{\rm BH}$ seems to confirm the hypothesis
that the 2 enigmatic sources are located in \ngc. Finally, and perhaps
most importantly, our analysis suggest the presence of 2 IMBHs within
a few hundred parsecs from the center of the galaxy, which hosts a
small SMBH. This result, if confirmed, may have important implications
for our understanding of the formation process of supermassive black
holes. Indeed, several theoretical models predict the existence of
off-nuclear supermassive black holes, which eventually merge with the
central SMBH (see, e.g., Volonteri \& Madau 2008 and reference
therein).

\section{Conclusions}

We have analyzed a 25 ks \chandra\ observation of the bulgeless galaxy
\ngc\ to investigate the basic X-ray properties of the putative AGN
and of 2 bright X-ray sources detected $\sim$20\arcsec\ away from the
center.  The main results can be summarized as follows:

\begin{itemize}

\item A weak X-ray source was detected at the center of the galaxy,
lending support to the finding from IR and optical spectroscopic
studies that \ngc\ hosts an AGN. Unfortunately, the signal-to-noise
ratio hampers any spectral study of the source. We therefore used
PIMMS to infer the possible luminosity associated with the putative
AGN. Assuming an absorbed power-law model with $\Gamma=1.7-2$ and
$N_{\rm H}=5\times10^{21} - 5\times10^{23}$ \nh, the resulting
luminosity is $L_{\rm 0.5-8\, keV}\sim 7\times 10^{37} - 3\times
10^{39}$ \lum\ ($L_{\rm 2-10\, keV}\sim 5\times 10^{37} - 2\times
10^{39}$ \lum), depending on the choice of intrinsic absorption.

\item Combining the X-ray luminosity with information derived from
recent IR and optical observations, and using several independent
tests, we infer that a heavily absorbed AGN scenario is more likely
for \ngc. If this is the case, exploiting the direct correlation
between bolometric correction factor and Eddington ratio, we speculate
that the AGN is in its early phase with a supermassive black hole of
relatively small mass ($M_{\rm BH}\simeq 2\times 10^4 {\rm~ M_\odot}$)
accreting at high rate ($L_{\rm bol}/L_{\rm Edd}\ga 0.2$).

\item The \chandra\ observation also reveals the presence of 2 bright
off-nuclear sources located $\sim$20\arcsec\ away from the central
source. Direct fitting of the X-ray spectra indicates that both are
well described by absorbed power laws with luminosities of the order
of $10^{39}$ \lum, in the hypothesis that they are located in \ngc.

\item The lack of strong optical/UV counterparts and the 
intrinsic absorption
argue against the 2 off-nuclear sources being Galactic objects that
appear to be in \ngc\ by chance superposition. Similarly, the
combination of X-ray and optical measurements seems to rule out the
hypothesis of background AGN.
Assuming that these
sources are at the distance of \ngc, leads to the conclusion that
their luminosities exceed the Eddington limit for a neutron star
and suggests that these are black holes. 

\item In order to discriminate between 2 competing scenarios for these
putative black hole systems (namely, 
IMBHs, or bright microquasars similar to GRS~1915+105 in \ngc), we
have applied 2 different methods that allow to determine the $M_{\rm
BH}$-to-distance ratio, once the spectrum is fitted with the
\verb+BMC+ model Comptonization model. Both methods consistently
suggest that the 2 off-nuclear sources are located in \ngc\ and have
$M_{\rm BH}$ of a few thousand solar masses.

\end{itemize}

In conclusion, our \chandra\ observation has confirmed that \ngc\ is
an extremely interesting object. Not only do the X-ray data support
the scenario that \ngc\ hosts a buried AGN, but they also suggest the
possible presence of 2 IMBHs located relatively close to the center of
the galaxy.  These findings may play a crucial role in our
understanding of the connection between galaxy formation and SMBHs at
their center, and of the process leading to the formation of the SMBH
itself. It must be, however, kept in mind that these results were
derived from data with limited statistics. Therefore, deeper X-ray
exposures from more sensitive instruments complemented with
observations at longer wavelengths (with emphasis on the radio band
that is largely unaffected by absorption) will be necessary to confirm
and possibly further our main conclusions.

\begin{acknowledgements} 
We thank the referee for the constructive 
comments and suggestions that improved the clarity of the paper.
MG acknowledges support by the \chandra\ Guest Investigator Program
under NASA grant GO8-9112X.

\end{acknowledgements}

\begin{table}[ht]
\caption{Possible optical counterparts of off-nuclear X-ray sources}
\begin{center}
\scriptsize
\begin{tabular}{ccccccc}
\hline
\hline
\noalign{\smallskip}
 Source & RA  & DEC         &  Offset & $F_{\nu}$ (F435W)      & $F_{\nu}$ (F555W)            & $F_{\nu}$ (F814W) \\
\noalign{\smallskip}
     &(J2000)   & (J2000)  &  ('')   & ($\mu$Jy) & ($\mu$Jy)    &    ($\mu$Jy) \\
\noalign{\smallskip}       
\hline
\noalign{\smallskip}
B    &   11:18:15.180  &  -32:48:40.39  &  0.33  &  1.8 (2.5) $\pm$ 0.4 & 1.4 (1.8) $\pm$ 0.5 &  $<$4.5 ($<$5.2) \\
\noalign{\smallskip}
C1   &   11:18:18.248   & -32:48:53.29 &   0.37 & 0.52 (0.72) $\pm$ 0.06& 0.46 (0.59) $\pm$ 0.06& 0.74 (0.85) $\pm$ 0.16\\
\noalign{\smallskip}
C2   &   11:18:18.252  &  -32:48:52.56  &  0.52 & $<$0.19 ($<$0.26) &$<$0.17 ($<$ 0.22)& 0.52 (0.60) $\pm$ 0.16\\
\noalign{\smallskip}
\hline
\end{tabular}
\end{center}
\footnotesize
The limits of the monochromatic fluxes are 3 sigma. The pivot wavelengths of the filters 
are respectively
4318 \AA\ (for F435W), 5360  \AA\ (for F555W), and 8060 \AA\ (for F814W). Extinction 
corrected (based on Schlegel et al. 1998 maps) values are indicated in parentheses.
\label{tab1}
\end{table}

\end{document}